\documentclass[11pt]{elsart}
\usepackage[utf8]{inputenc}
\usepackage[IL2]{fontenc}
\usepackage[english]{babel}
\usepackage{amsmath}
\usepackage{graphicx}
\usepackage{float}
\usepackage{xcolor}
\usepackage[version=4]{mhchem}
\usepackage{amssymb}
\usepackage{amsmath}
\usepackage{amsfonts} 
\usepackage{siunitx}

\newcommand{\Cerenkov}{$\rm \check{C}$erenkov $\,$}

\begin{document}
\begin{frontmatter}

\title{Consequences of the CMR effect on EELS in TEM}

\author[ustem]{Wolfgang Wallisch}
\author[ustem]{, Michael Stöger-Pollach}
\author[cta]{and Edvinas Navickas}

\address[ustem]{University Service Centre for Transmission Electron Microscopy, Technische Universität Wien, Wiedner Hauptstraße 8-10, A-1040 Wien, Austria}
\address[cta]{Institute of Chemical Technologies and Analytics, Technische Universität Wien, Getreidemarkt 9, A-1040 Wien, Austria}

\date{\today}
\begin{abstract}
Double perovskite oxides have gained in importance and exhibit negative magnetoresistance, which is known as colossal magnetoresistance (CMR) effect. Using a \ce{La2CoMnO6} (LCM) thin film we prove that the physical consequences of the CMR effect do also influence the electron energy loss spectrometry (EELS) signal. We observe a change of the band gap at low energy losses and are able to study the magnetisation with chemical sensitivity by employing energy loss magnetic chiral dichroism (EMCD) below the Curie temperature T\textsubscript{C} where the CMR effect becomes significant.

\end{abstract}


\end{frontmatter}

\section{Introduction}
Over the last decades double perovskite manganites have been of great interest in many fields such as cathodes of solid oxide fuel cells (SOFC) for high temperature applications \cite{Navickas2015}, magnetic storage and magnetic field sensing functions at room temperature \cite{Haghiri-Gosnet2000}. Their magnetic \cite{Dass2003}, electric \cite{Masud2012}, dielectric \cite{Silva2016}, magnetoelectric \cite{Yang2014} and magnetoresistance \cite{Mahato2010} properties are promising for a wide variety of applications. In particular, the colossal magnetoresistance (CMR) of perovskite oxides is an interesting metal-insulator transition describing the change of the resistance in the presence of a magnetic field \cite{Haghiri-Gosnet2003}.

Transmission electron microscopes (TEMs) equipped with energy filters are powerful tools and are routinely used for chemical and microstructural analysis of small regions. The main advantage is the high spatial resolution for probing band gaps, which is substantial limited by the inelastic delocalization of the respective energy loss and the excitation of \Cerenkov light caused by the high velocity of the probe electrons. Moreover, high beam energies may destroy some materials, whereas a decrease of the beam energy reduces this problem. Lowering the speed of the probe electron achieves a reduction of the inelastic delocalization and avoids the excitation of \Cerenkov losses, too. As a further consequence, the zero loss peak (ZLP) tails are narrowed leading to an improvement of the analysis of the band structure as well as the dielectric properties. This is a very positive effect for semiconductor and insulator analysis \cite{Stoger-Pollach2010}. 

Fundamental material properties caused by the valence and conduction bands, like plasmon resonances, band gaps and optical properties are probed in the low loss part of the EELS spectrum covering approximately the first 50~eV. Focusing on the transitions from the valence band into excited states is in general called Valence EELS (VEELS). Compared to optical methods it allows a larger energy range (1~eV to 50~eV) and angular dependent measurements, which means that the momentum transfer is not equal zero. Since conventional optical methods are limited by the wave length of light in the determination of band gaps and optical properties, VEELS has become an accurate solution not at least because of its high spatial resolution. An additional advantage of TEM is the opportunity to detect electron loss magnetic circular dichroism (EMCD) of few nanometres that enables to investigate the magnetisation of the specimen with a high spatial resolution \cite{Schattschneider2006,Schattschneider2008} and chemical sensitivity \cite{Ennen2012}.

The resistivity and the magnetic field of double perovskite \ce{La2CoMnO6} (LCM) manganites are showing different behaviour below the Curie temperature T\textsubscript{C}, although a low magnetic field is merely applied \cite{Mahato2010,KrishnaMurthy2012,Viswanathan2010}. The investigations of \textit{Mahato et al.} \cite{Mahato2010} and \textit{Y$\acute{a}\tilde{n}$ez-Vilar et al.} \cite{Yanez-Vilar2009} showed a difference of the magnetisation below 210~K if a field of 100~Oe is applied or not. In addition \textit{Mahato et al.} \cite{Mahato2010} reported that the relation between the magnetisation and the applied field under consideration of the temperature dependence exhibits at 10~K a coercive field of $\sim$6~kOe, while the coercive field at room temperature (RT) is almost zero. The temperature dependence of the resistivity was investigated inter alia by \textit{Y$\acute{a}\tilde{n}$ez-Vilar et al.} \cite{Yanez-Vilar2009} and \textit{Yang et al.} \cite{Yang2014}. Furthermore the value of the magnetoresistance (MR) is decreasing if the temperature is increasing and a CMR of 80~\% is achieved at 5~K by applying a magnetic field of 80~kOe \cite{Mahato2010}.
For this purpose, LCM oxides are well suited for proving consequences of the CMR effect on EMCD and VEELS depending on magnetisation, resistivity and temperature, respectively. 

We therefore present experiments varying first the magnetic field of the objective lens magnetising the LCM film and secondly the temperature of the specimen.


\section{Experimental Procedure}
The LCM thin film samples were synthesised by pulsed laser deposition (PLD) technique. The stoichiometric compound PLD target was prepared from LCM powder. The 80~nm LCM layer was epitaxially grown on (001)-oriented \ce{SrTiO3} (STO) substrate at $760\,^{\circ}{\rm C}$ with a background pressure of 0.12~mbar in 20~min. The cross section lamella of the specimen was prepared by focused ion beam (FEI Quanta 200 3D DualBeam-FIB) followed by soft \ce{Ar+} ion polishing at 1~keV. 

All EELS experiments were performed using conventional TEMs (FEI TECNAI) equipped with a thermionic electron gun system (\ce{LaB6}) and a GATAN post-column energy filter (GIF 2001), and one with a field emission gun and a GATAN Tridiem energy filter. The TEMs can be operated in the free high-tension mode and enable the free choice of the acceleration voltage in the range from 10~kV to 200~kV. The cooling of the sample was done by employing a cryo-transfer holder (GATAN 915) reaching a minimum temperature of 85~K.

The band gaps and optical properties are detected in the low loss part of the energy loss spectrum by VEELS. During the measurements at different temperatures the relative thickness at the position of interest was approximately 0.3~$\lambda$, which means that $\lambda$ is the mean path length for inelastic electron scattering at the respective beam energies. These were set to 40~kV and 200~kV. The spectrometer dispersion setting was 0.1~eV per channel and the collection angle was 8.4~mrad. The good balance of the signal to noise ratio was achieved by multiple VEELS spectrum acquisitions. For the 200~keV experiments we summed over 50~spectra with recording times of 0.2~s, whereas for the 40~keV experiments we summed over 30~spectra with recording times of 1.6~s. The final energy resolution was 0.9~eV in both cases.

\begin{figure}[ht]
	\centering
	\includegraphics[width=0.5\textwidth]{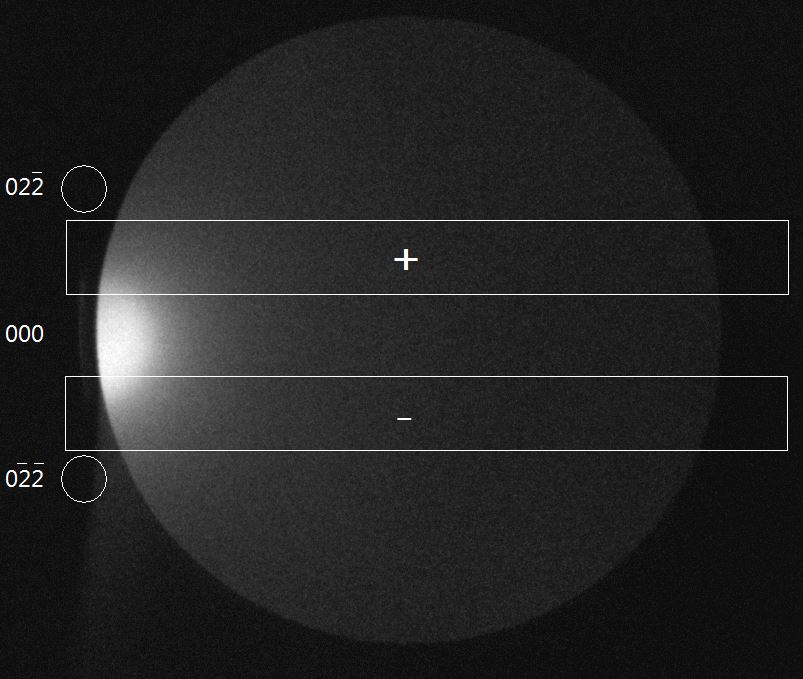}
\caption{Experimental three-beam case (3bc) set-up for the EMCD experiments. The definition of the chiral positions ``+'' and ``-'' is as given in the diffraction pattern.}
\label{Fig1}
\end{figure}

EMCD allows to characterise the magnetisation of the observed material in the nano scale range and with chemical sensitivity. The corresponding experiments were performed in the classical scheme similar to the one suggested in \cite{Schattschneider2008a} using the crystal as the beam splitter to achieve specific three-beam diffraction geometries. This set-up guarantees a superposition of coherent wave vectors at the chiral positions defined in Figure \ref{Fig1}. 

Using a slightly convergent 200~keV electron beam (convergence angle $<$ 0.5~mrad) guarantees only to illuminate 60~nm of the 80~nm wide LCM layer. The observed sample region had an approximate thickness (in the direction of the electron beam) of 20~nm to 25~nm. The EMCD signal then results from the difference of the EELS spectra on the ''+'' and ''-'' position. The temperature and the magnetic field within the objective lens of our TEM can be varied. Therefore, the magnetic field applied to the specimen can be changed and generates different observations.


\section{Results and discussion}

In order to study the influence of the CMR effect onto the EELS spectrum, we are varying the magnetic field and the temperature by changing the beam energy and cooling the specimen, respectively. When we reduce the beam energy from 200~keV to 40~keV the objective lens field is adjusted from 1.9~T to 1.2~T at the sample position. Thus the CMR effect will increase the band gap width and hence reduce the conductivity of LCM. At 85~K this change is within the experimental limits in terms of energy resolution. At room temperature we are above  T\textsubscript{C}~=~210~K, thus no change in conductivity is expected. Additionally, the EMCD effect will only be measurable below T\textsubscript{C}. Consequently we divide this chapter into a low-loss section dealing with the band gap and interband transitions, and a core-loss section discussing the EMCD results.

\subsection{Low losses}
\begin{figure}[ht]
	\centering
\includegraphics[width=1\textwidth]{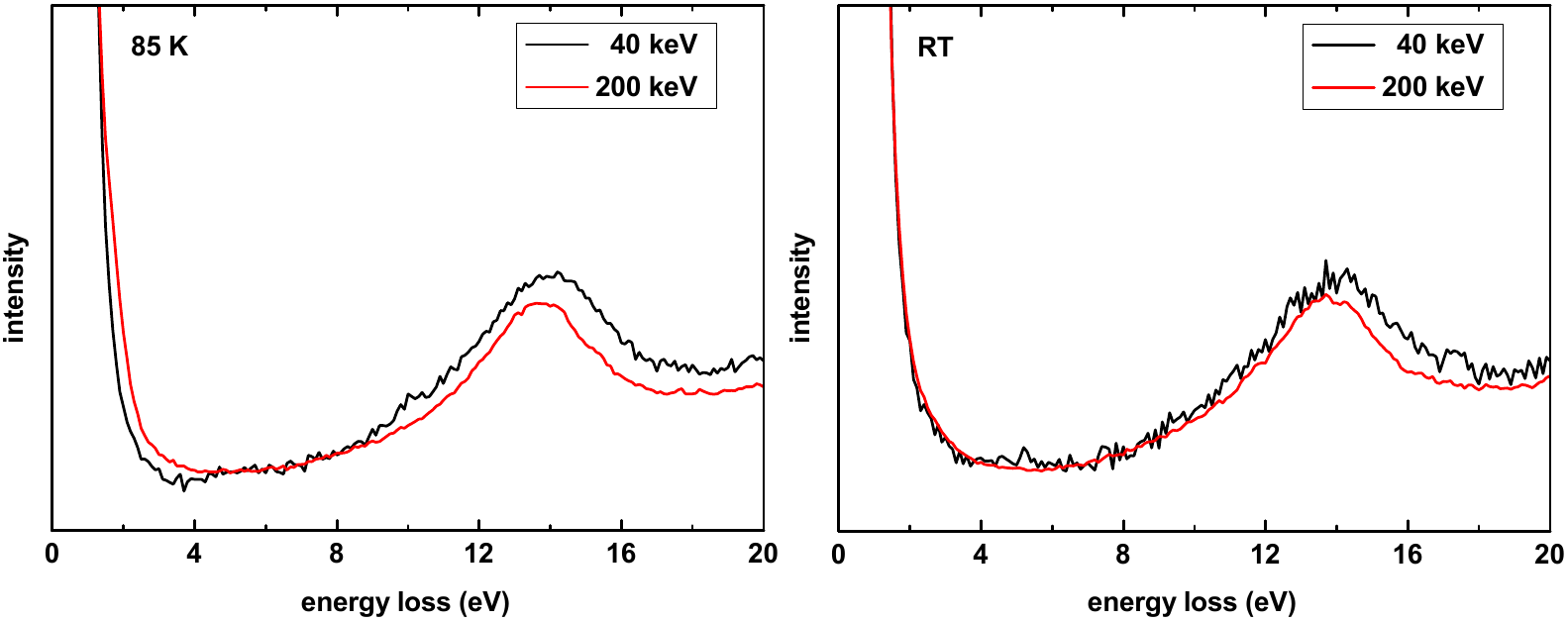}
		\caption{Left: unprocessed VEELS spectra recorded at 85~K at a sample thickness of 0.3~$\lambda$ using 40~keV and 200~keV, respectively.  Right: unprocessed VEELS spectra recorded at RT at a sample thickness of 0.3~$\lambda$ using 40~keV and 200~keV, respectively.}
\label{Fig2}
\end{figure}

\begin{figure}[ht]
	\centering
\includegraphics[width=0.5\textwidth]{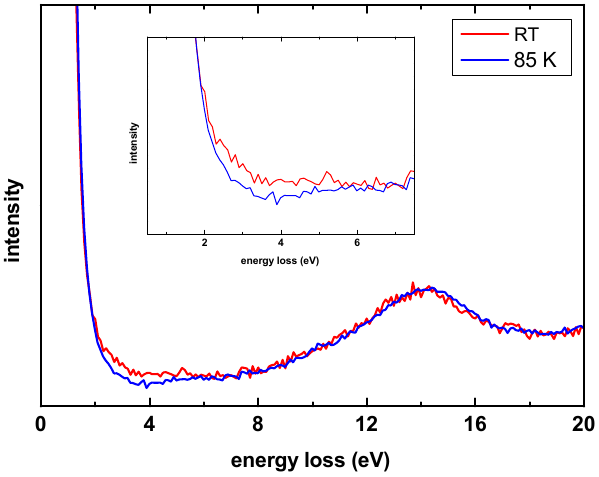}
		\caption{VEELS spectrum recorded at RT and 85~K using 40~keV electrons. The insertion shows the divergences between the RT and the 85~K spectrum.}
\label{Fig3}
\end{figure}

\begin{figure}[ht]
	\centering
\includegraphics[width=0.5\textwidth]{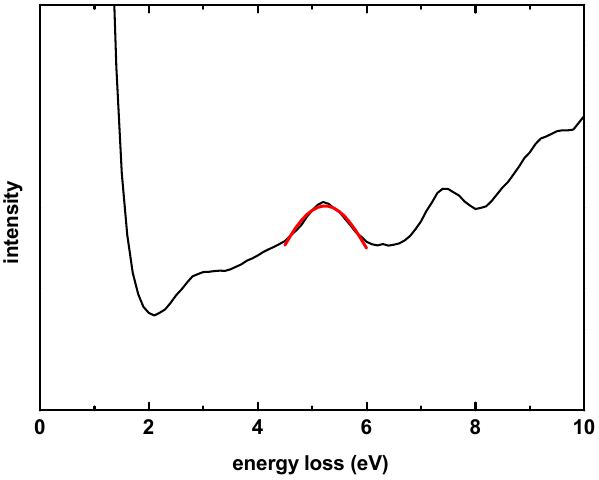}
		\caption{RL smoothed spectrum at 40~kV and RT.}
\label{Fig4}
\end{figure}

The low loss spectrum in EELS contains information about the dielectric response of the specimen to the perturbation caused by the probe electron. Hence interband transitions and band gaps can in principle be determined. Nevertheless, the \Cerenkov effect has to be considered, which means that fast electrons excite radiation when passing through a medium being faster than the light would be.
$$
v_e = \frac{c_{0}}{n}
$$
where $v_e$ is the speed of the probe electron, $c_0$ is the vacuum speed of light and $n$ is the refractive index. Due to the fact that the refractive index of the investigated material is $n = 2.4$, as being determined by means of Kramers-Kronig Analysis of the 40~keV experiments, the \Cerenkov limit of the beam energy is approximately 50~keV \cite{Horak2015}. Figure \ref{Fig2} shows the low loss spectrum in the range from 0~eV to 20~eV energy loss. The left hand side gives the spectra at 85~K and the right hand side shows the ones at RT. It is obvious that there is in both cases a difference in the energy range of 1.5~eV to 4~eV energy loss. It can be either caused by the \Cerenkov effect, which adds intensity due to the excitation of the \Cerenkov radiation in the 200~keV spectra or it can be caused by the CMR effect, which narrows the band gap and thus adds intensity in the very low loss region of the VEELS spectra. In order to minimise the influences of the \Cerenkov effect on the VEELS spectra there are two possibilities: (i) probing an extremely thin sample area \cite{Stoger-Pollach2006} or (ii) to reduce the beam energy. Basically we do both, because of varying the magnetic field of the objective lens at the sample area is needed in order to see the gap narrowing. 
On the other hand, when comparing the 40~keV results (shown in Figure \ref{Fig3}) at 85~K with the ones recorded at RT, there is no \Cerenkov effect influencing the spectra. Thus the intensity variation is caused by the CMR effect only. Additionally, the smaller magnetic field enhances the variation of the band gap width with respect to temperature.
This is in agreement to \cite{Mahato2010}, the resistivity of LCM at an applied magnetic field decreases with increasing temperature. 
The direct determination of the band gap is not easy, because of the high-energy tails of the zero loss peak (ZLP) \cite{Stoger-Pollach2008}. Therefore we measured the shift of the strongest interband transitions being visible in the VEELS spectrum as small peaks at 2.9~eV, 5.3~eV and 7.6~eV energy loss. For this purpose we used the Richardson-Lucy algorithm \cite{Gloter2003} for deconvolving the spread of the electron source. Figure \ref{Fig4} shows the 40~keV VEELS spectrum recorded at RT after 5 iterations. The interband transitions are clearly visible and can easily be fitted by a Gaussian for the determination of its exact position using a non-linear least square fit routine.
For proving the stability of this method the fitting range was shifted around the interband transition by several channels of the spectrum. In the case of the RT experiment at 40~keV the interband transition was determined to be at 5.240~eV with standard deviation of 0.017~eV. The interband transition for the investigation at 85~K was found at 5.384~eV with a standard deviation of 0.084~eV. Consequently an energy band shift of 0.14~eV for the measurements at RT and 85~K exists at an external magnetic field of 1.2~T. The results for both beam energies and both temperatures are summarised in Table \ref{Tab1}.\\
\begin{table}[h!]
\begin{tabular}{l|cc}
                 & 85 K & RT \\
\hline
200 keV & 4.92 & 4.88 \\
40 keV  & 5.38 & 5.24 \\
\end{tabular}
\caption{The values of two different interband transitions fitted by a Gaussian for both beam energies (40 keV and 200 keV) and both temperatures (85 K and RT).}
\label{Tab1}
\end{table}
%


\subsection{EMCD}
\begin{figure}[ht]
	\centering
\includegraphics[width=0.9\textwidth]{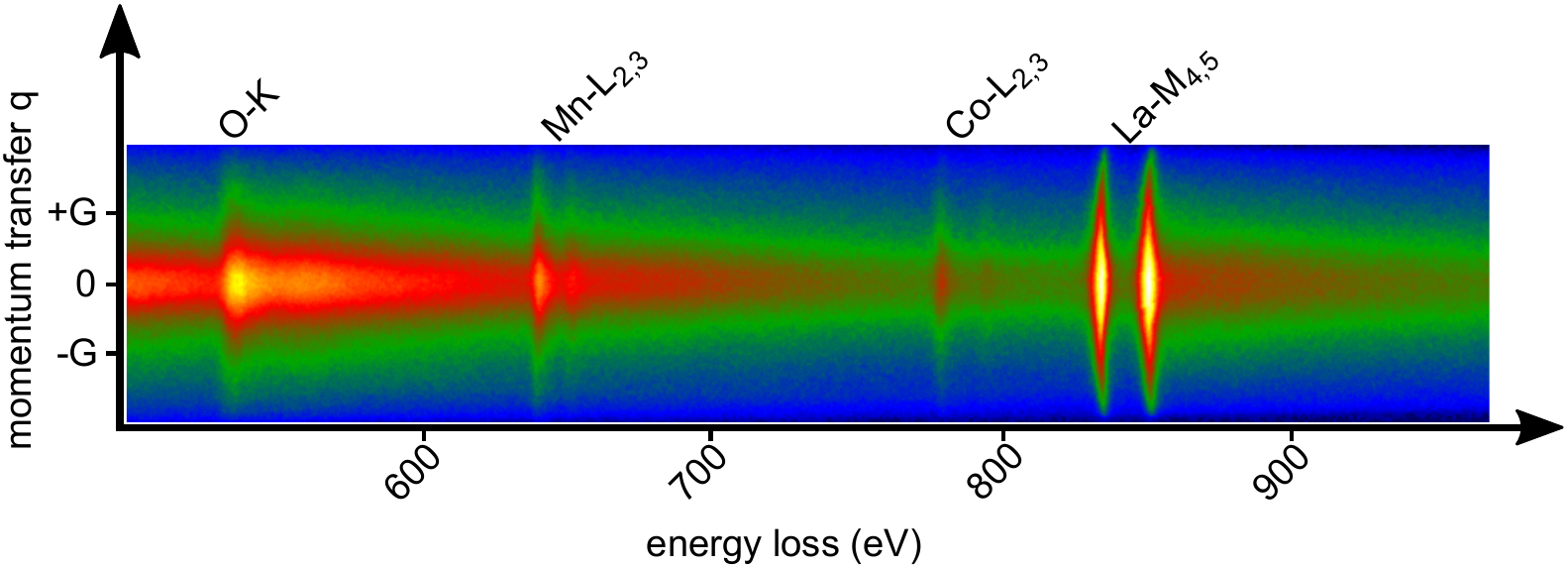}
		\caption{Two-dimensional EELS spectrum of the LCM layer showing the edges of the corresponding elements. G is defined as (220).}
\label{Fig5}
\end{figure}

On the basis of several investigations \cite{Silva2016,Mahato2010,Yang2014}, which propose a temperature dependent magnetic behaviour, we study the influences of the CMR effect on EMCD and we vary the sample's temperature only. A variation of the external magnetic field in the range from 1.9~T to 1.2~T (200~keV beam energy to 40~keV beam energy) would not give any influence on the EMCD signal, because in both cases the magnetic moments are fully saturated and aligned with respect to the objective lens field. Consequently, EMCD experiments are performed only at 200~keV. Figure \ref{Fig5} shows the $\Delta$E-$\vec{q}$ map including the O-K edge, the Mn-L$_{2,3}$ edges, the Co-L$_{2,3}$ edges and the La-M$_{4,5}$ edges. The EMCD effect can be determined by the asymmetric angular distribution of the respective energy losses, when integrating over the Co-L$_3$ edge, only. Following \cite{Schattschneider2008a} we calculate the angular distribution of the EMCD signal in Figure \ref{Fig5}.
\begin{figure}[ht]
	\centering
\includegraphics[width=0.5\textwidth]{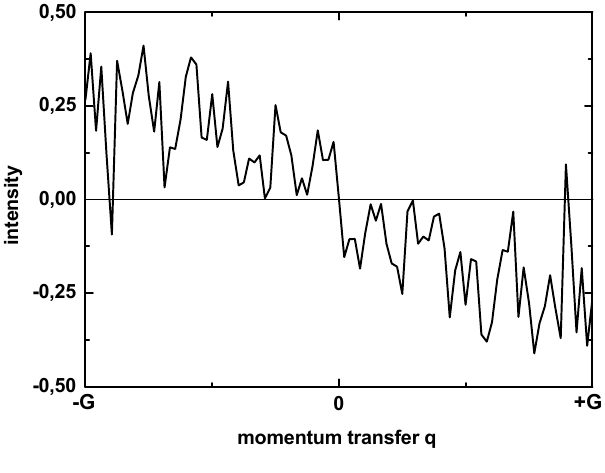}
		\caption{Dichroic signal at the cobalt L$_3$ edge as a function of the scattering angle $\vec{q}$ (in unit of G) in the direction perpendicular to the Bragg scattering vector $\bf{G}$, which is (220).}
\label{Fig6}
\end{figure}

\begin{figure}[ht]
	\centering
\includegraphics[width=0.8\textwidth]{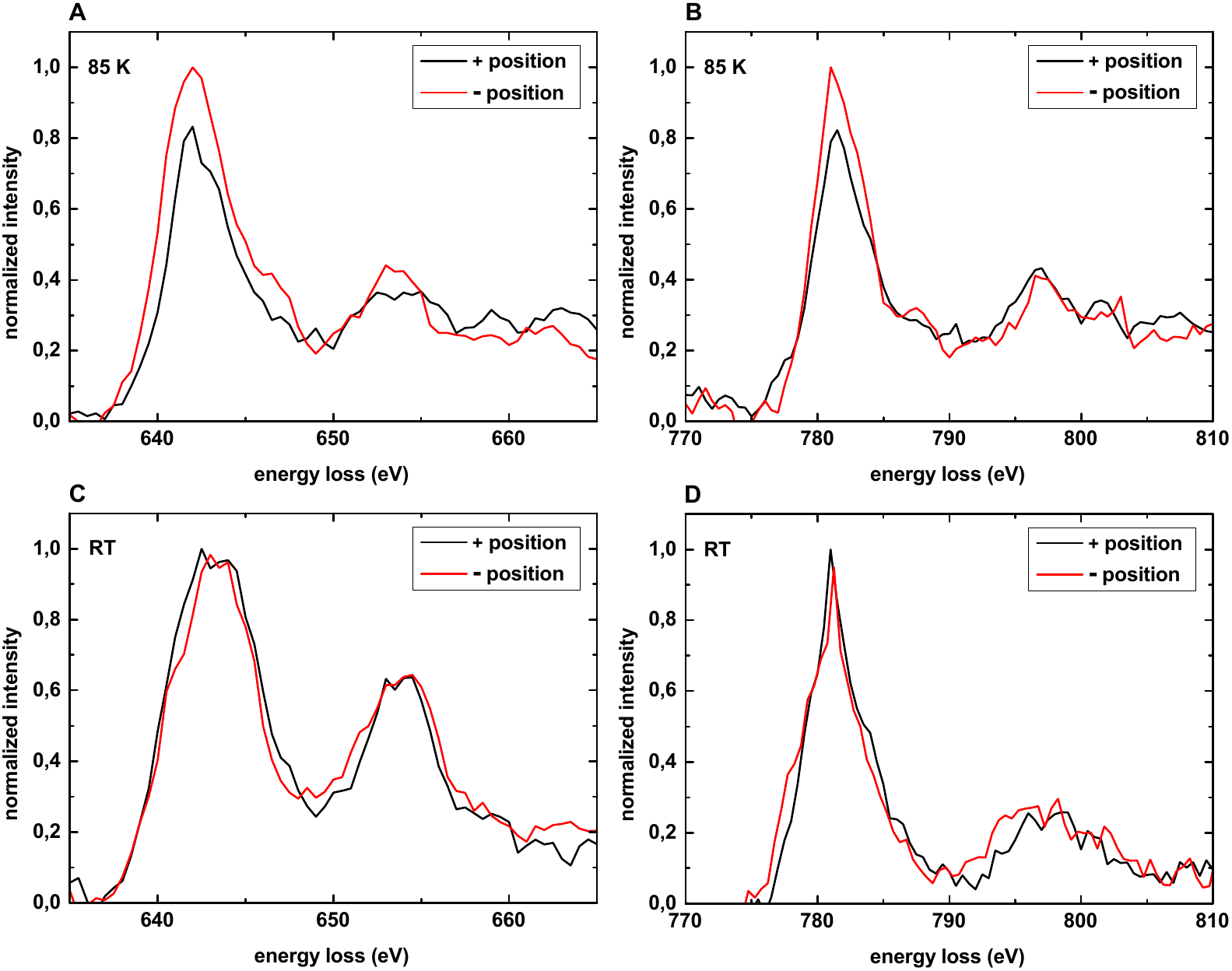}
		\caption{Normalized EELS spectra of the LCM layer. The (A) Mn and the (B) Co edges show induced chiral electronic transitions at 85~K. (C) and (D) show the same edges at RT, where no EMCD effect can be observed.}
\label{Fig7}
\end{figure}

The chemical sensitivity of EMCD is demonstrated in Figure \ref{Fig7}. In the upper row the L$_{2,3}$ edges of Mn (640~eV energy loss) and Co (779~eV energy loss) are presented from the measurement at 85~K. The lower row of Figure \ref{Fig7} shows the respective results from the RT experiment. At 85~K the CMR effect causes a magnetisability in a magnetic field of less than approximately 0.5~T \cite{Mahato2010}, the objective lens field is strong enough to fully magnetise the Mn and Co atoms. Consequently the EMCD effect can be observed in the 85~K experiment. In contrast, at room temperature the lens field is not strong enough and thus no EMCD signal can be measured.
The EMCD signal strength was not simulated with respect to the sample thickness \cite{Rubino2008}. The optimum EMCD sample thickness might be not used during the experiments. Hence, the magnetism of the LCM thin film was proved at 85~K and confirmed the expected results with an effective EMCD signal of about 5~\% at the Mn (Figure \ref{Fig7}a) and Co (Figure \ref{Fig7}b) edges. A quantitative statement about the magnetisation cannot be given due to the not optimized sample thickness and the noisy data, caused by short recording times due to sample drift. 


\section {Conclusion and Outlook}
By observing the double perovskite oxide LCM we could detect the influences of the CMR effect on the band gap and the element specific magnetisation by means of EELS. Even though the energy resolution during the experiments was in the order of 0.9~eV, a shift of the band gap by 0.14~eV can be measured by using the interband transition losses. Additionally the magnetisation was proven by employing EMCD. For this purposes we varied the magnetic field of the objective lens in the vicinity of the specimen and changed the sample's temperature. Consequently EELS can be employed at least qualitatively to investigate electronic properties of  CMR-materials in both the states above and below T\textsubscript{C}. Improving the sample stability will lead to a better signal-to-noise ratio in both, the VEELS and the EMCD spectra. Thus the obtained results will be of a quantitative nature, since the EMCD spectra can in principle be analysed for the magnetic moments in terms of orbital moment and spin state. Using a monochomator will certainly improve the VEELS investigations, hence giving access to more exact  interband transition strengths and energies.


\section*{Acknowledgements}
The authors kindly acknowledge financial support by the Austrian Science Fund (FWF; F 4501 and F 4509).



\bibliography{biblio}

\end{document}